# Influence of directional spreading on wave overtopping of sea dikes with gentle and shallow foreshores

CORRADO ALTOMARE[1,2], TOMOHIRO SUZUKI[3,4], TOON VERWAEST[3]

[1] *Universitat Politecnica de Catalunya – BarcelonaTech (UPC), Spain, corrado.altomare@upc.edu*

[2] *Ghent University, Belgium, corrado.altomare@ugent.be*

[3] *Flanders Hydraulics Research, Belgium, toon.verwaest@mow.vlaanderen.be, tomohiro.suzuki@mow.vlaanderen.be*

[4] *Faculty of Civil Engineering and Geosciences, Delft University of Technology, Stevinweg 1, 2628 CN Delft, The Netherlands*

**ABSTRACT**

The work highlights the importance of directional spreading effects on wave overtopping estimation in shallow and mild sloping foreshores. Wave short-crestedness leads, in general, to a reduction of mean overtopping discharges on coastal structures. In the present work, the case of a sea dike with gentle foreshore in very and extremely shallow water conditions is analysed. Physical model tests have been carried out in order to investigate the effect of directional spreading on overtopping and incident wave characteristics. In the present experimental campaign, the effect of wave spreading has only been investigated for perpendicular wave attack. Results show that directional spreading is proved to cause a reduction of average discharge of sea dikes with gentle and shallow foreshore. Expressions for the reduction factor for directional spreading are derived, fitted on the tested database. The use of this reduction factor leads to more accurate prediction and avoids overtopping overestimation, however reduction-factor formulations are overtopping-formula depending.

**KEYWORDS:** Wave overtopping; directional spreading; reduction factor; shallow foreshore; sea dike.

## 1. INTRODUCTION

Coastal areas worldwide are at risk because of anthropogenic and natural hazards, which are expected to increase due to changing climate (Weisse *et al.*, 2012). Effects of the climate change, such as the sea level rise and the occurrence of more severe and frequent storms represent major threats to coastal defences. The mean sea level has been increasing during the last century on average of 1-2 mm/year, tendency that is worsening during the last few decades (Jevrejeva *et al.*, 2006). Besides, the 20[th] century has seen already a number of severe storms causing damage and flooding worldwide. Examples of these storms are: the North Sea flood of 1953, considered to be the worst natural disaster of the 20th century both in the Netherlands, Belgium and the United Kingdom, claiming 2,551 lives and leading to damages for more than 0.6 bn USD; Xynthia in 2010 (63 casualties and damages for more than 1.4 bn USD); Xaver in 2013 (15 casualties, ≥ 1.3 bn USD). Worldwide, the Hurricane Katrina caused over 125 bn USD damages and 1,800 casualties only in USA in 2005, probably the most destructive hurricane in the latest 20 years in USA, followed by Harvey (2017, 68 fatalities and 125 bn USD damages) and Sandy (2012, 233 fatalities 68.7 bn USD). In the same geographical area, in 2017 the Hurricane Maria struck and devastated Dominica, the U.S. Virgin Islands, and Puerto Rico, with over 3,000 fatalities and 91.61 bn USD damages. All the aforementioned events are just a few examples of a long list of severe weather conditions that are likely to occur again, enhanced by the climate change. In particular, low-lying countries are ones of the most exposed areas to wave overtopping and sea flood. These countries are characterised by densely populated and low-elevation coastal areas that, despite the increasing risk for flooding, are experiencing a continuous population growth. In many low-elevation coastal areas, very shallow, long and gentle foreshores lie in front of the coastal protections. Only a few studies are available in literature on wave overtopping prediction for such a beach layout and specifically in combination with sea dikes (van Gent *et al.*, 2007; Altomare *et al.*, 2016; Suzuki *et al.*, 2017). These studies analysed the case of long-crested wave conditions, being based on wave flume experimental campaigns or 2DV numerical modelling. However, it is of high importance to understand the influence of gentle and shallow foreshores for real three-dimensional sea states (short-crested waves) on wave transformation and wave overtopping. Guza & Feddersen (2012) demonstrated influence of directional spreading for wave run-up and Suzuki et al. (2014) showed one for wave overtopping by phase resolving wave models. However, those were limited to the numerical modelling. While numerical solvers can help to characterise wave transformation and overtopping for short-crested waves (Zijlema *et al.*, 2011; Roelvink *et al.*, 2009), not many data are available for a proper model validation are available for the aforementioned conditions with shallow and gentle foreshores under realistic sea states, i.e. short-crested waves. Physical model tests have been usually carried out for structures lying on horizontal bottom and deep or intermediate water conditions ate the toe, not taking into account the influence of gentle and shallow foreshores (e.g. Nørgaard et al., 2014; van Gent and Van der Werf, 2019). Besides, the behaviour of free and bound infragravity waves over a sloping bottom (Janssen *et al.*, 2003; Battjes *et al.*, 2004; van Dongeren *et al.*, 2007) under realistic sea states will be of interest. It would be important to take into account the characteristics of free and bound long waves, which dominate the hydrodynamics in the shallow foreshore, in order to understand overtopping phenomena in shallow foreshore condition better.

Physical model tests have been carried out in the shallow-water wave basin at Flanders Hydraulics Research (FHR) in Antwerp, Belgium, to analyse the influence of directional spreading on wave overtopping and post-overtopping processes on sloping sea dikes with 1:35 foreshore slope in case of very and extremely shallow water conditions (Hofland et al., 2017). The experimental campaign is part of the CREST (Climate REsilience coast) project (http://www.crestproject.be/en), a Belgian-funded project and the goal of which is to increase the knowledge of coastal processes nearshore and landward.

In the present work, the influence of wave short-crestedness on mean overtopping discharge is discussed. The results are compared with existing semi-empirical formulae from literature. This work aims at representing a first step towards a more comprehensive understanding of wave overtopping of sea dikes for cases with very and extremely shallow foreshores due to real three-dimensional sea states.

## 2. SHALLOW WATER CRITERIA

The foreshore can be defined as the part of the seabed bathymetry in front of the dike toe, that causes processes like wave breaking and refraction. The most recent criterion to define the shallowness of the foreshores has been published by Hofland *et al.* (2017). The authors characterise the shallowness of the foreshore by means of the ratio of the still water depth near the structure, $h_t$, by the offshore wave height, $H_{m0,o}$, in deep waters (Figure 1). Table 1 shows the ranges of foreshore shallowness as in Hofland *et al.* (2017). These criteria allow avoiding misinterpretation of the shallowness as for cases with non-breaking swells on deep foreshores. In the present research, cases from very shallow to extremely shallow foreshores, which imply heavy wave breaking, have been modelled.

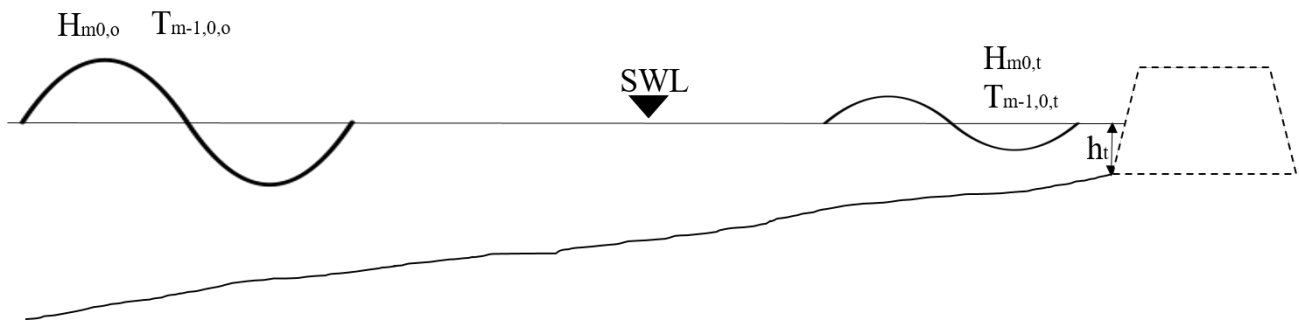

*Figure 1. Layout for the definition of offshore and toe conditions. The offshore conditions are indicated by the offshore spectral wave period $T_{m-1,0,o}$ and the offshore wave height $H_{m0,o}$. The conditions at the toe of a possible structure are indicated by the spectral wave period, $T_{m-1,0,t}$ the wave height at the toe $H_{m0,t}$ and the water depth at the toe and $h_t$. Still water level (SWL) is indicated, as well.*

Table 1. Ranges of foreshore shallowness (Hofland *et al.*, 2017)

| Deep | $\frac{h_t}{H_{m0,o}} > 4$ |
|---|---|
| Shallow | $1 < \frac{h_t}{H_{m0,o}} < 4$ |
| Very Shallow | $0.3 < \frac{h_t}{H_{m0,o}} < 1$ |
| Extremely Shallow | $\frac{h_t}{H_{m0,o}} < 0.3$ |

A new set of formulas for the wave period calculation at the toe of the dike were proposed by Hofland *et al.* (2017). For long-crested waves, the increase of the spectral period from offshore to the toe of the dike is expressed by:

$$\frac{T_{m-1,0,t}}{T_{m-1,0,o}} - 1 = 6\exp(-4\tilde{h}) + \exp(-\tilde{h}) \tag{1}$$

where $T_{m-1,0,o}$ and $T_{m-1,0,t}$ are the spectral wave period offshore and at the toe respectively and a $\tilde{h}$ is new parameter for relative water depth defined as follows:

$$\tilde{h} = \frac{h_t}{H_{m0,o}}\left(\frac{cot\theta}{100}\right)^{0.2} \tag{2}$$

being *cotθ* the foreshore slope, ranging between 35 and 250 in Hofland *et al.* (2017). For short-crested waves, a formula with similar shape was presented:

$$\frac{T_{m-1,0,t}}{T_{m-1,0,o}} - 1 = 6\exp(-6\tilde{h}) + 0.25\exp(-0.75\tilde{h}) \tag{3}$$

## 3. AVERAGE WAVE OVERTOPPING ASSESSMENT IN EXISTING LITERATURE

Wave overtopping occurs when sea waves run up coastal defences which are not high enough to prevent flows over their crest. Wave overtopping is a very complex phenomenon because it varies in time and space during the same storm event. It is common practice to assess the mean or average overtopping discharge that can be defined as the ratio between the total volume of water that overtops a coastal defence by the duration of the storm event. Hence, mean wave overtopping is widely used worldwide as one of the most important design criteria for coastal defences. It is a simple but meaningful parameter. Mean overtopping depends not only on the local wave conditions, but also on the dike geometry (crest elevation, presence of storm walls or blocks, etc.). In areas characterized by gentle and shallow foreshores, the foreshore slope affects also the overtopping discharge (Altomare et al., 2016), because of the provoked heavy breaking and propagation of the resulting broken waves on the same foreshore.

Average overtopping can be estimated by means of numerical or experimental modelling or by employing semi-empirical formulas, which rely and are fitted on experimental results or in situ measurements. This section offers a brief review of those semi-empirical formulas that might be applicable to the case of wave overtopping on sea dikes with very or extremely shallow foreshores.

### 3.1 Goda (2009)

Goda (2009) proposed a set of unified formulas for the prediction of the mean rate of wave overtopping of smooth impermeable coastal structures both for sloping and vertical structures. For that scope, he analysed a specific dataset derived from the database of the EU-funded CLASH project (in total 1254 data), considering only cases with smooth, impermeable faces. Excluded from the calibration datasets were also the data of vertical walls with re-curved wave walls or broad crests, and the data with oblique wave incidence (angle of wave attack not being 0°). Goda used also 198 tests carried out at Kansai University, Japan, by (Tamada *et al.* 2001) that presented dike slopes of 1:3, 1:5 and 1:7 and foreshore slopes equal to 1:10 and 1:30. No wave conditions at the toe were provided in this latter dataset, reason why Goda calculated the predicted wave height at the toe by means of the method presented in (Goda, 2006).In Goda (2009), the author showed that existing exponential formulas have a tendency to overestimate large overtopping rates and underestimate low overtopping rates, when calibrated with the extracted dataset. He used the simple well known exponential formula for wave overtopping:

$$\frac{q}{\sqrt{gH_{m0,t}^3}} = Q = \exp\left[-\left(A + B\frac{R_c}{H_{m0,t}}\right)\right] \tag{4}$$

where *q* is the mean overtopping discharge expressed in m³/s/m, $R_c$ is the freeboard, $H_{m0,t}$ is the incident spectral wave height at the dike toe, *g* is the gravity acceleration. The author redefined the coefficients A and B not just as constant values, but as functions of the dike slope, the foreshore slope and the dimensionless toe depth.

$$A = A_0 tanh[(0.956 + 4.44 tan\theta) \cdot (h_t/H_{m0,t} + 1.242 - 2.032 tan^{0.25}\theta)] \tag{5}$$

$$B = B_0 tanh[(0.822 - 2.22 tan\theta) \cdot (h_t/H_{m0,t} + 0.578 + 2.22 tan\theta)] \quad (6)$$

with:

$$A_0 = 3.4 - 0.734 cot\alpha + 0.239 cot^2\alpha - 0.0162 cot^3\alpha \quad (7)$$
$$B_0 = 2.3 - 0.5 cot\alpha + 0.15 cot^2\alpha - 0.011 cot^3\alpha \quad (8)$$

being $\theta$ the foreshore angle with the horizontal, $h_t$ the still water depth at the dike toe and $\alpha$ the dike slope angle. The Eqs.(5-8) are valid in the range 0≤cot α≤7. Both A and B coefficients increase up to a constant value if the relative toe depth increase. The plateau is reached for $h_t/H_{m0,t}$ bigger than 3.0 in both cases. However, all collected data are characterized by a dimensionless toe depth $h_t/H_{m0,t}$ bigger than 1.0, therefore lacking of data for very and extremely shallow water conditions.

### 3.2 Mase *et al.* (2013)

Mase et al. (2013) proposed a set of formulae for sea dikes with very shallow foreshore and even emergent toe. Their formulae are based on deep-water wave characteristics. They first calculate the expected run-up and then define an imaginary slope used for overtopping calculation. The imaginary slope is introduced to overcome the difficulties in schematizing complex dike geometries. Wave breaking depth is also an important parameter for the application of their method. Uncertainties in the calculation of the breaking depth and aspects like neglecting the wave directional spreading occurring between offshore location and the dike toe, make the use of their formulae of restricted application.

### 3.3 Van Gent (1999) modified by Altomare *et al.* (2016)

Altomare et al. (2016) introduced the concept of equivalent slope in shallow water conditions. The equivalent slope is applied to re-assess the surf-similarity parameter, which will be used in the overtopping formula that keeps the original structure as proposed by van Gent (1999). The influence of water depth at the toe, foreshore slope and dike slope are considered by using the equivalent slope. The authors suggest to use the equivalent slope when the ratio $h_t/H_{m0,t}$ is smaller than 1.5, otherwise the dike slope only should be considered for further calculations. Hence, it is assumed that the foreshore starts to influence the wave overtopping when the toe depth is smaller than 1.5 times the incident wave height. The authors use data from CLASH database and data obtained at experimental campaigns at Flanders Hydraulics Research and Ghent University (in total 279 data with shallow or very shallow foreshore conditions). The relative toe depth ranges between -0.25 m and 3.65 m. The new equivalent slope concept can be applied also to cases of dry toe (= negative water depth at the toe). Mean wave overtopping can be assessed using the following equation:

$$\frac{q}{\sqrt{gH_{m0,t}^3}} = Q = 10^c \exp\left[-\frac{R_c}{H_{m0,t}(0.33 + 0.022\xi_{m-1,0})}\right] \quad (9)$$

The *c* exponent in Eq. (9) is assumed to be normally distributed: under the hypothesis the mean value of *c* results equal to −0.791 and the standard deviation σ is 0.294. The dike or structural slope to calculate the surf similarity parameter, $\xi_{m-1,0}$, has to be replaced if $h_t/H_{m0,t} \leq 1.5$. In such a case, the average slope between the point on the foreshore with a depth of $1.5H_{m0,t}$ and the run-up level $R_{u2\%}$, can be expressed as follows:

$$tan\alpha_{sf} = \frac{(1.5H_{m0,t} + R_{u2\%})}{(1.5H_{m0,t} - h_t) \cdot cot\theta + (h_t + R_{u2\%}) \cdot cot\alpha} \quad (10)$$

For the calculation, both structure and foreshore are assumed to have a straight slope without a berm, defined by *cotα* and *cotθ* respectively. Therefore, the expression of $\xi_{m-1,0}$ becomes:

$$\xi_{m-1,0} = \frac{tan\alpha_{sf}}{\sqrt{s_{m-1,0}}} \quad (11)$$

being $s_{m-1,0}$ the wave steepness.

## 3.4 EurOtop (2018)

While being already a worldwide reference for wave overtopping assessment, EurOtop (2018) provides limited information on wave overtopping for very shallow water cases because of the lack of extensive databases on such cases. For very shallow water cases, the equation of Altomare *et al.* (2016) is reported in (EurOtop, 2018). For very steep slopes up to vertical walls a new formula is proposed in EurOtop (2018), in which the coefficients of the formula depend on the dike slope. For slopes steeper than 1:2 (V:H) there is an influence of the slope angle, for more gentle slope there is no influence based. This formula is applied to vertical walls only when there is no influence of the foreshore. Therefore, the dike slope might play an important role, leading to conclusions similar to the ones from Goda (2009). Nevertheless, the formula does not cover cases with very shallow waters.

## 4 EXPERIMENTAL CAMPAIGN

Physical model experiments have been carried out in the wave basin equipped with multi-directional wave generation system at FHR. Within the framework of the CREST project, the wave basin was employed primarily to study the effects of wave overtopping and post-overtopping processes (e.g. overtopping wave layer characteristics and force) of the short-crestedness of the waves in very and extremely shallow water conditions with the presence of a gentle foreshore.

### 4.1 Wave basin setup and generation system

The wave basin at Flanders Hydraulics Research is 17.9 m wide and 23.2 m long (Figure 3) having a T-shape where the two side zones are conceived to allow shore-parallel current generation and to place damping material as passive absorption system. The effective model area in front of the wave generator is 12 x 20 m. The maximum operating depth is 0.55 m. The basin is equipped with a multi-directional wave generation system, comprising 30 piston paddles (each paddle 0.4 m wide) with electric actuators. The independent movement of the paddles allows both short-crested and oblique waves to be generated. The maximum paddle stroke is 1.1 m. The system has been built and installed by HR Wallingford, together with the wave generation software HR Merlin, which embeds a reflection compensation system. Resistive wave gauges placed on each piston paddle are employed to measure the free surface and correct the paddle movement in order to absorb the reflected wave components. The maximum regular wave height that can be generated is 0.25 m, 0.13 m in case of significant wave height for random sea states. The extended basin method (Dalrymple, 1989) can be used instead of the more conventional method to generate oblique long-crested waves. Three different wave spectra can be used, namely JONSWAP, Pierson-Moskowitz and TMA. A user defined spectrum can be specified as well. The new system also allows generating solitary waves and focused wave groups.

The physical model tests consisted in 3D experiments with a fixed bed. A 1:35 foreshore slope was built in concrete, representing an average (eroded) profile along the Belgian coast. The foreshore starts at 5.35 m far from the position of the wave paddles at rest and extends for 8.89 m. Before the foreshore a 1:10 (V:H) transition slope was built. At the end of the foreshore, a 1:2 (V:H) slope sea dike is located (Figure 2). The axis origin is at the wave paddle position at rest. Both dike and promenade are built in high density polyurethane. The dike height, measured vertically from the toe to its seaward edge is 0.05 m in the model scale. The model is split into two different study areas after the dike, see Figure 3: on the right, overtopping boxes are installed right after the dike crest; on the left a 1:50 (V:H) promenade of 0.4 m width is installed. At the end of the promenade a vertical wall, made of PVC, is installed, modelling the façade of buildings built along the coastline.

Detailed measurements of wave propagation and transformation on the foreshore and overtopping with impact loading on the buildings on top of the sea dike have been gathered during the experiments. However, those data are not part of the present analysis. Instead, we focus only on average overtopping discharges. It was thereby verified the negligible influence of the asymmetric layout in the basin on the incident wave characteristics at the toe of the dike (see next section).

In total 56 instruments have been employed. The ones used for the present study comprise: one star-array consisting in 7 wave gauges to measure the incident offshore wave field and wave directionality; seven wave gauges located at the toe of the dike to measure the wave characteristics at that location (WG$_{toe}$ 1-7 in Figure 3); three stainless steel overtopping tanks to measure average and individual overtopping equipped with load cells and one Baluff magnetostrictive linear position sensor. Load cells, four per overtopping tank, measure the variation in weight due to the water overtopping and flowing into the tank. Baluff transducers measure the distance between a position magnet, which is floating with the free surface and the head end of the sensing rod: knowing the area of tank is then possible to calculate the wave-by-wave overtopping volume. Usually employing load cells avoid noise typical of the Baluff measurement for overtopping and that must be filtered out, which derives from high frequency water level oscillations within the tank.

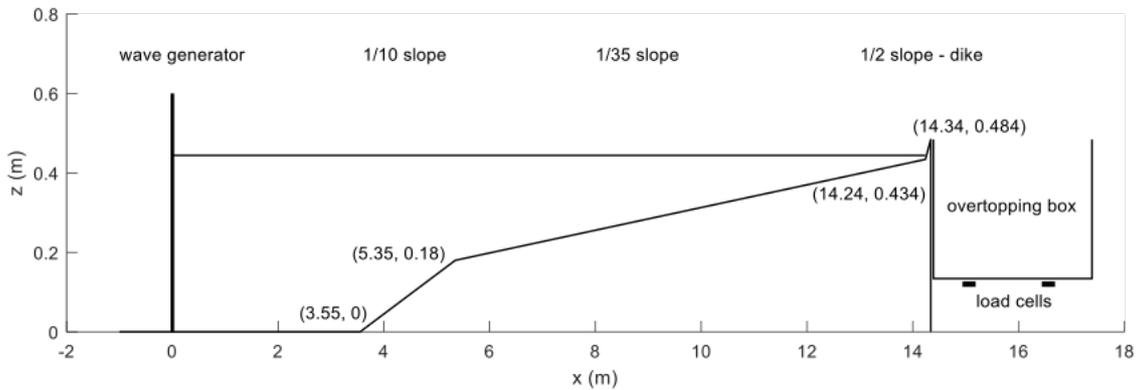

*Figure 2. Cross section of foreshore and dike profile. Horizontal dimensions are distorted.*

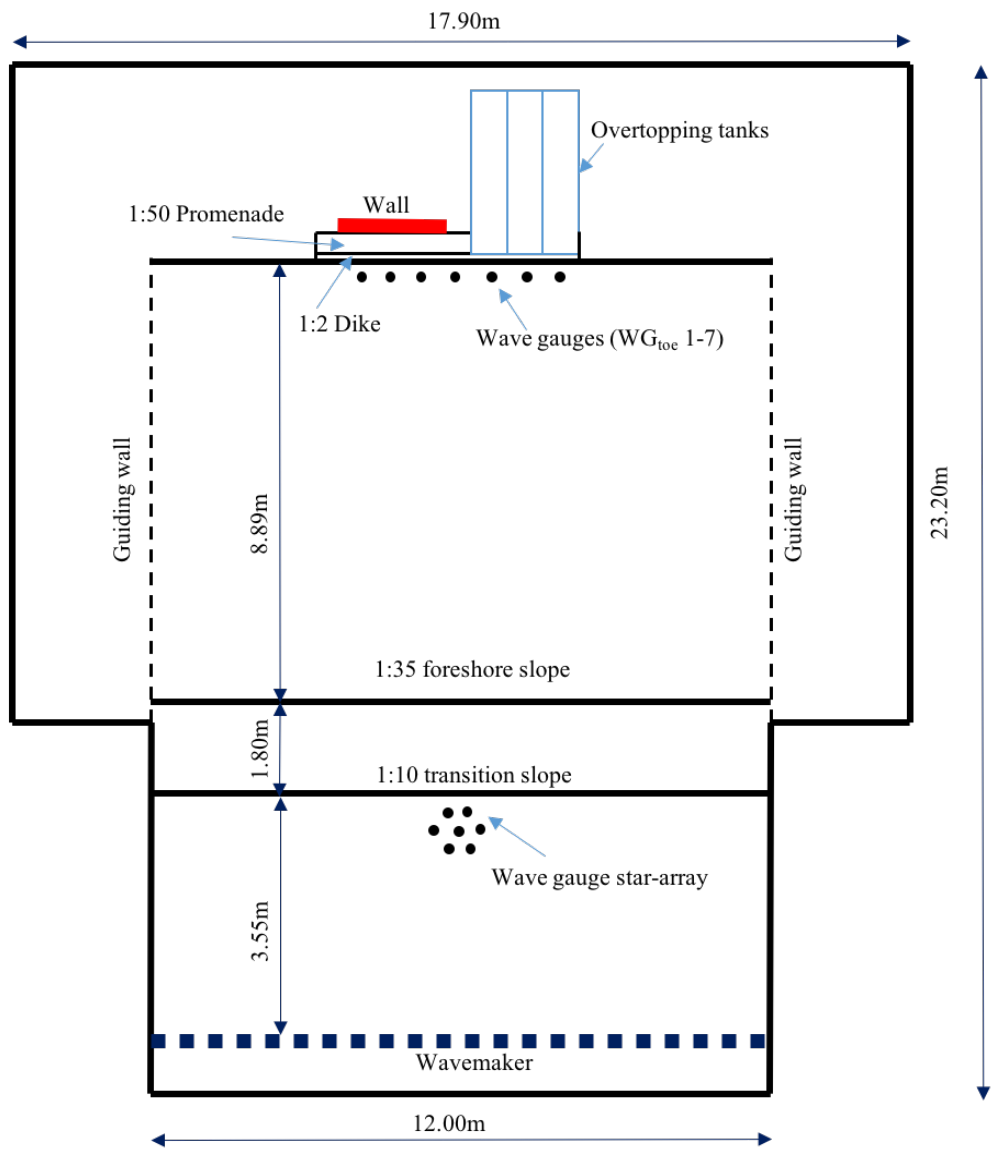

*Figure 3. Plan view of the experimental layout (not in scale).*

## 4.2 Wave conditions

Froude similarity is applied to scale down the prototype conditions and layout. The employed model scale was 1:50. Hereafter all wave conditions and results will be expressed in prototype scale (1:1). Stormy wave conditions, having a return period of 1,000 and 17,000 years *(Verwaest et al., 2008)*, have been chosen to be reproduced in the wave basin. Target offshore wave heights of 3 m, 4 m and 5 m with peak periods of 10 s and 12 s have been modelled. The offshore water level varies between +7 m TAW and +8 m TAW in prototype, which correspond to a water depth of 22.1 m and 23.1 m respectively. TAW stands for Tweede Algemene Waterpassing (Second General Levelling in Dutch), which is the Belgian standard datum level. The still water depth at the toe of the dike is between 0.5 m and 1.5 m. The seaward edge of the dike crest is at +9 m TAW which correspond to a freeboard between 1 m and 2 m.

The wave conditions being generated are summarised in Table 2. The wave-generation software did not allow combining directional spreading and obliqueness. Therefore, the main direction of short-crested waves was perpendicular to the dike. In total 125 tests were carried out. Forty-six tests employed long-crested wave waves, 26 of which with perpendicular wave propagation (obliqueness equal to 0°). Test cases with long-crested oblique waves (23 in total) are not analysed in the present work.

Table 2. Offshore target wave characteristics in prototype scale.

| Water level [m] | $H_{m0,o}$ [m] | $T_{p,o}$ [s] | Spreading [°] |
|---|---|---|---|
| 22.1 (+7 m TAW) | 3.0 | 10 | 0, 12, 16, 20, 31.5 |
| 23.1 (+8 m TAW) | 4.0 | 12 | |
| | 5.0 | | |

Repeatability has been checked in terms of local wave conditions at the toe of the dike and mean overtopping discharges. Two different wave conditions, corresponding to significant wave height offshore of 3 m and 5 m and for +7 m TAW water level have been repeated three times. The calculated coefficient of variation σ', defined as the ratio between standard deviation σ and mean value μ, has been calculated, resulting equal to 1%, 2% and 5% for wave height, period and overtopping respectively, within acceptable ranges (EurOtop, 2018).

A preliminary check on the influence of wave generation theory on wave transformation and overtopping has been carried out. Second-order wave generation (namely including correction for bound long waves) has been employed to generate and absorb long-crested waves. However, the generation of second-order short-crested waves was not available in the used version of HR Merlin software and almost all cases of short crested waves have been generated employing first-order theory. In a second phase, time series of piston displacement including bound long-wave correction have been generated externally for the case corresponding to $H_{m0,o}$=3 m, $T_{p,o}$=10 s and directional spreading of 12° and 16° with water level equal to +7 m TAW. These wave conditions have been selected based on the expected overtopping discharge, namely around 0.5-1 l/s/m. The time series have been used by the generation software to generate second-order short-crested waves. It is worthy to notice that, in this specific case, accurate absorption of long-wave components is not guaranteed. Five different seed numbers have been used for each wave conditions. The coefficient of variation (σ' =σ/μ) of wave height, period and overtopping has been estimated based on the different seeding and resulted in σ'($H_{m0,t}$)=1-3%, σ'($T_{m-1,0,t}$)=3-4%, σ'(q)=19-30%, respectively. The differences due to wave order generation theory have been quantified. For long-crested waves, a reduction in wave height of 5% was measured using a second-order wave theory. Variation in period and overtopping discharges were around 6%. Small differences at the toe level were measured (±5%), being in the order of the accuracy of the magnetostrictive linear position sensor, which might lead to differences in wave transformations and, hence, on wave overtopping, especially for wave conditions characterised by the presence of a saturated surf zone, as highlighted Franco *et al.* (2009). For short-crested waves, no significant variation of local conditions was measured between first- and second-order wave theory. The lack of a proper wave generation and absorption system for second-order short-crested waves could mask the influence of directional spreading. Nevertheless, the differences in overtopping using different generation theories in the specific case above fall within the uncertainties related to the seed number as described in Williams *et al.* (2019), where variability of the overtopping discharge was found between 20% and 75%. Hence, despite the model limitations due to the generation software, the effect of the wave order for further analysis was neglected.

After executing all tests and having acquired all data, the model (including dike, walls and overtopping tanks) has been removed and only the foreshore slope has been left. A horizontal bottom followed by passive absorption material have been placed on the rear side of the foreshore slope. Such a configuration guarantees to damp the wave reflection and allows measuring the incident wave conditions at the location corresponding to the dike toe. Note that the well-known reflection analysis methods (e.g. Mansard and Funke) are not fully applicable to the condition at the toe of the dike in this campaign since the waves are highly non-linear. Due to the non-uniform spatial distribution of wave height and period, the average of the spectral wave height and mean period over all 3 gauges in front of overtopping tanks (WG$_{toe}$ 5-7) is used for mean wave overtopping assessment. The spatial variability of local conditions is mostly due to small construction uncertainties of the foreshore, which result in slightly different water depth at the toe, and other model effects (such as reflection from the later guiding walls). The coefficient of variation has been employed to quantify the uncertainty due to spatial distribution, resulting σ'($H_{m0,t}$)=0.6-6.6% and σ'($T_{m-1,0,t}$)=0.3-4%. The influence of it has been quantified in terms of mean overtopping discharge by means of Eq. (9), showing variation of q of at most 1.7 times, largely within the uncertainties of the formula prediction. The range of conditions at the dike toe is shown in Table 3.

Table 3. Measured overtopping and incident waves characteristics at the position of the dike toe expressed in prototype scale.

| Water depth at the dike toe [m] | $H_{m0,t}$ [m] | $T_{m-1,0,t}$ [s] | q [l/s/m] |
|---|---|---|---|
| 0.46-1.53 | 0.61-1.51 | 19.96- 62.47 | 0.25-173.7 |

### 4.3 Scale effects

The employed small model scale could result in significant scale effects on wave overtopping due to surface tension and viscous forces in some cases. The influence of viscous forces and surface tension has been analysed by calculating the Reynolds and Weber number for wave overtopping ($Re_q$ and $We_q$) respectively and by comparing the results with the critical limits, namely $Re_q>10^3$ and $We_q>10$. For further details, see also EurOtop (2018). Only 5 tests did not satisfy the condition $Re_q>10^3$. The Weber number as calculated for all tests varies between 25 and 39, bigger than the critical limit of 10. The 5 tests with $Re_q<10^3$ have been excluded from further analysis, resulting in a database of 97 tests with negligible scale effects for viscous forces and surface tension.

The tested water levels and wave heights at the structure toe are relatively small (of ca. 1cm), following typical guidelines on wave run-up and overtopping modelling. To exclude model effects e.g. the water depth and wave height and need to be in the correct ranges, as recommended in EurOtop (2018) and Frostick et al. (2011): water depth h>5cm and wave height $H_{m0}$>2cm. Since this is not the case here where extremely shallow waters lead to very small values of water depth and incident wave height at the dike toe in model scale, further model effects cannot be completely ruled out. Hence, it is recommended, whenever possible, to validate the results at a larger scale if a quantification at prototype scale of average discharge is pursued for such structural layout and for design purposes.

## 5  RESULTS AND DISCUSSIONS

### 5.1 Average overtopping discharge

The results in terms of average wave overtopping discharge are summarized and discussed in the present section. First a general overview of all results, including long-crested and short-crested waves is given. Later on, a more detailed analysis of the influence of short-crestedness is carried out. Finally, the results are compared versus the prediction of Eq. (9) and Eq. (4). The measured average discharge (expressed in l/s/m) is plotted against the incident wave height at the toe in Figure 4. Bigger the wave height, larger the overtopping discharge, however differences can be already noticed between long-crested cases (diamond markers), which gives the larger overtopping discharges, and short—crested test cases (circle markers).

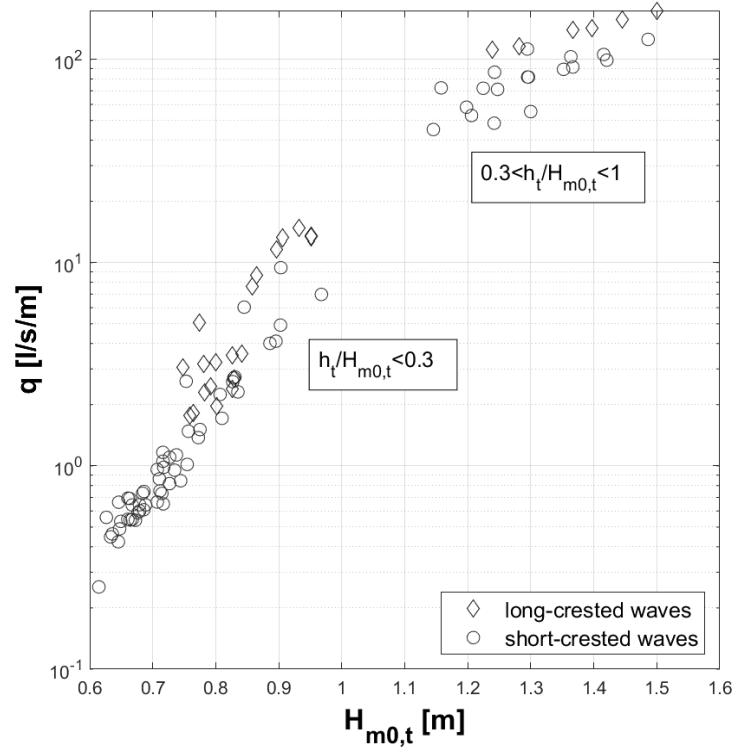

*Figure 4. Average overtopping discharge versus incident spectral wave height at the dike toe (prototype conditions). Data are split in two groups: very shallow water conditions (0.3<$H_{m0,t}$<1) and extremely shallow water conditions ($H_{m0,t}$<0.3).*

The data plotted gather in two groups corresponding respectively to very (0.3<$h_t/H_{m0,t}$<1.0) and extremely ($h_t/H_{m0,t}$<0.3) shallow water conditions. The former group corresponds to initial still water level of +8 m TAW and toe depth of 1.5 m. The latter one is characterised by +7 m TAW and toe depth of 0.5 m. Mean overtopping discharge for $h_t/H_{m0,t}$<0.3 is one or more order of magnitude smaller than the one measured for 0.3<$h_t/H_{m0,t}$<1.0. Main reason is the increase of the dimensionless freeboard, $R_c/H_{m0,t}$, due to both the increase of the dike crest freeboard $R_c$ and reduction of the incident wave height at the toe caused by heavier wave breaking.

## 5.2 Shallowness regime

As previously mentioned, the wave conditions at the toe have been defined as very or extremely shallow. A first check on the application of Eq. (1) and Eq. (3) was made based on the surf-similarity-like parameter, $\beta_b$ defined as follows:

$$\beta_b = \theta T_{m-1,0,o} \sqrt{\frac{g}{H_{m0,o}}} \qquad (12)$$

Hofland's equations can be applied if $\beta_b$ <0.62. This threshold characterises mild slopes where shoaling of bound long waves is the dominant generation mechanism for infragravity waves. The value of $\beta_b$ for each test case is depicted in Figure 5 versus the relative water depth at the toe, $h_t/H_{m0,o}$. While all data are characterized by mild slope and $\beta_b$ is always minor than 0.62, the data are still gathered in two groups corresponding to very and extremely shallow water conditions.

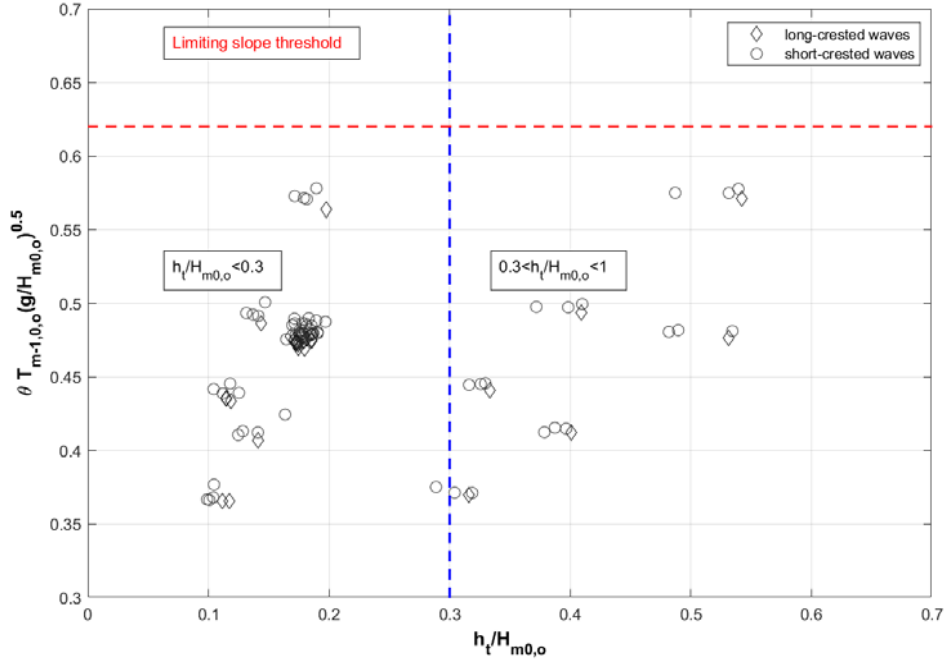

*Figure 5. Surf-similarity-like parameter function of relative water depth at the toe.*

The importance of water levels on overtopping results are clarified in Figure 6 where the variation of the discharge with the relative toe depth, $h_t/H_{m0,o}$ is depicted. As for Figure 4, the data are clearly split in two groups: very shallow water cases ($0.3<h_t/H_{m0,t}<1.0$) and extremely shallow water cases ($h_t/H_{m0,t}<0.3$). This distinction is very important since the existing formulas will perform differently for the two regimes, as described in the following sections. Within a regime overtopping decreases when $h_t/H_{m0,o}$ increases, however the influence is bigger for extremely shallow water conditions, where the discharge can differ of some order of magnitude, suggesting a different behaviour for different water depths to be further investigated (not part of the present research). Therefore, the water depth at the toe is playing an important role, as also anticipated by Goda (2009).

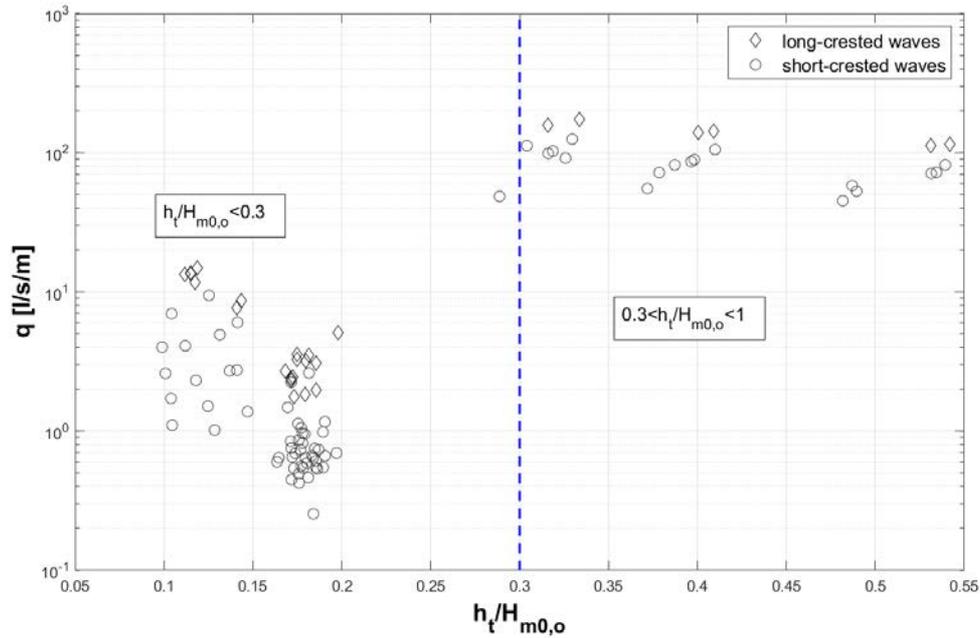

*Figure 6. Dependence of discharge on the ratio $h_t/H_{m0,o}$*

Finally the measured spectral wave period at the toe has been compared with predictions from Hofland *et al.* (2017). The period evolution from offshore to the toe of the dike has been compared with Eq. (1) and Eq. (3).

The results are depicted in Figure 7. Both plots of long-crested and short-crested waves follow the predictions and fall within the confidence interval.

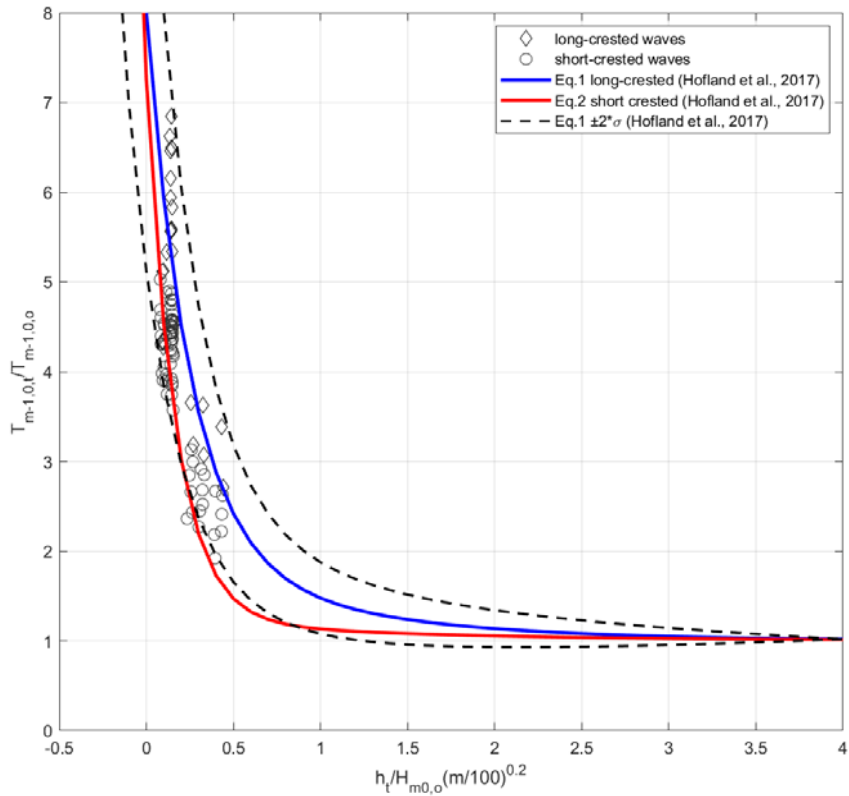

*Figure 7. Evolution of the spectral wave period as function of relative water depth and comparison with formulas from Hofland et al. (2017)*

### 5.3 Influence of the short-crestedness on overtopping and wave characteristics

The results of long-crested perpendicular and short-crested wave cases are compared in this section. It is expected that larger spreading values lead to lower overtopping discharges according to Guza & Feddersen (2012). The influence of the directional spreading is shown as ratio of short-crested values by corresponding long-crested wave cases (Figure 8). Wave height and period at the dike toe correspond to the average of $WG_{toe}$ 5-7. Overtopping discharge is averaged over the three wave tanks. The directional spreading measured in deep water is shown (using the star-array data). A clear distinction between the two different water levels can be noticed in the top and bottom plots of Figure 8: the influence of the spreading is bigger for lower water depth (+7 m TAW) than for higher water depth (+8 m TAW). The distinction between the two water levels is noticeable especially for wave height and average overtopping. The wave period reduces as well with the spreading, however it is not possible to distinguish clearly between higher and lower water levels. Values of the wave height and wave overtopping ration greater than 1 can be noticed in the figure, due to the fact that different seeding numbers have been used in some cases leading to different wave characteristics at the toe and different overtopping discharges. Even though the results are quite scattered, the influence of wave spreading can be seen clearly. The overtopping discharge reduces up to 1 order of magnitude for low water levels. The wave height decreases 10% and 20% in average for water levels equal to +8 m TAW and +7 m TAW, respectively. The maximum wave period reduction is about 35%. In general, the lower wave periods for short-crested waves than for long-crested waves confirm that the wave transformation and the mechanism leading to the release of infra-gravity waves act differently whether the spreading is present or not.

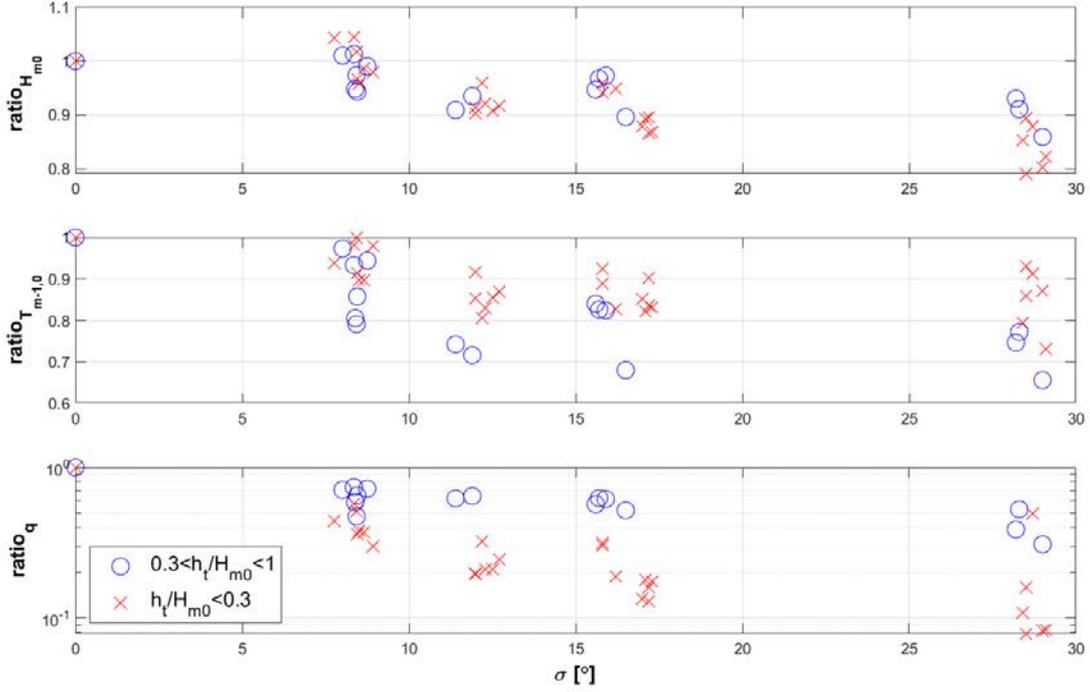

*Figure 8. Reduction of measured wave height (top), wave period (middle) and average overtopping (bottom) as function of the measured directional spreading*

The decrease of wave overtopping due to the wave short-crestedness has been assessed in terms of reductions factors. Besides directional spreading is proved to have on wave transformation and local wave characteristics at the toe of the dike, this is not enough to explain overtopping reduction. Directional spreading is expected to affect the whole process of waves running up the dike and overtopping it, if compared with similar incident wave conditions but from long-crested waves. The general way to calculate the value of the reduction factors is to assume it equal to ratio of mean discharge of the test where the reduction takes place with reference test, q/q$_{ref}$, however the requirement is that the wave conditions at the toe of the two tests must be the same (EurOtop, 2018). This condition was not met in this investigation. Hence, two different expressions of the reduction factor are proposed in the present work, derived for Eq. (9) and Eq. (4), respectively. The reader must consider that this approach has the limitation that each reduction factor formulation is formula depending. For Eq. (9), the reduction factor for each test has been calculated as follows:

$$\gamma_\sigma = \frac{Q_\sigma}{Q_\perp} \exp(R^*_\perp - R^*_\sigma) \qquad (13)$$

where $R^* = \left[\frac{R_c}{H_{m0,t,meas}(0.33+0.022\xi_{m-1,0,meas})}\right]$ and $Q = \frac{q_{meas}}{10^c\sqrt{gH^3_{m0,t,meas}}}$

The subscripts σ and ⊥ indicate the short-crested wave case and the reference long-crested wave case, respectively. Employing Eq. (4), the reduction factor is expressed as follows:

$$\gamma_\sigma = \frac{Q_\sigma}{Q_\perp} \exp[(A + BR)_\perp - (A + BR)_\sigma] \qquad (14)$$

where $R = \left[\frac{R_c}{H_{m0,t,meas}}\right]$ and $Q = \frac{q_{meas}}{\sqrt{gH^3_{m0,t,meas}}}$

The coefficients A and B are calculated based on Eqs. (5-8) for each test case. In this the difference between local condition at the toe is taken into account.

The results are plotted in Figure 9 together with the best fits which are expressed by the following equations.

For Altomare *et al.* (2016):

$$\gamma_\sigma = \begin{cases} \exp(-0.046\sigma), \text{ for } \frac{h_t}{H_{m0,o}} < 0.3 \\ \exp(-0.014\sigma), \text{ for } 0.3 < \frac{h_t}{H_{m0,o}} < 0.1 \end{cases} \quad (15)$$

For Goda (2009):

$$\gamma_\sigma = \begin{cases} \exp(-0.053\sigma), \text{ for } \frac{h_t}{H_{m0,o}} < 0.3 \\ \exp(-0.02\sigma), \text{ for } 0.3 < \frac{h_t}{H_{m0,o}} < 0.1 \end{cases} \quad (16)$$

The $R^2$ for each fit is indicated in the figures: low accuracy is shown only when Eq. (4) is applied for extremely shallow water conditions. This was expectable because of the lack of data in such conditions employed by Goda (2009) to derive the expressions of the A and B coefficients. Figure 9 confirms the clear distinction for $0.3<h_t/H_{m0,o}<1$ and $h_t/H_{m0,o}<0.3$, as shown previously is in Figure 8. Larger wave spreading leads to stronger reduction of wave overtopping, but this reduction is more pronounced for $h_t/H_{m0,o}<0.3$ than for $0.3<h_t/H_{m0,o}<1$, at least within the range of the tested conditions.

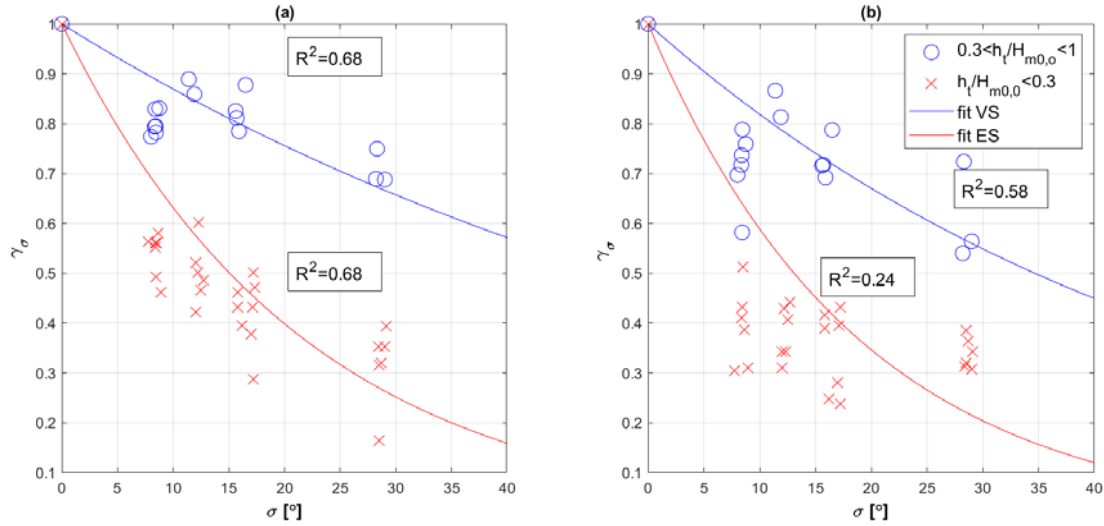

*Figure 9. Calculated reduction factors for wave short-crestedness employing Eq. (9) (a) and Eq.(4) (b), respectively. The solid lines indicate the best fit: red for extremely shallow waters (ES) and blue for very shallow waters (VS).*

**5.4 Average overtopping prediction**

The results of mean overtopping discharge are compared with predictions employing Eq. (4) and Eq. (9). Distinction is made between long-crested (depicted in figures as diamonds) and short-crested wave cases (circles). Both formulas take into account combined influence of the foreshore slope, dike slope and local water depth for overtopping assessment, however in different ways:

1. van Gent (1999) as modified by Altomare *et al.* (2016): equivalent or average slope is calculated assessing the wave run-up by iteration and used in Eq. (9).
2. Goda (2009): coefficients A and B are calculated by means of Eqs. (5-8) and used in Eq. (4) to assess mean overtopping discharge.

Reduction factors for directional spreading have been applied for overtopping prediction. The revised Eq.(9) can be expressed as follows:

$$\frac{q}{\sqrt{gH_{m0,t}^3}} = \gamma_\sigma 10^c \exp\left[-\frac{R_c}{H_{m0,t}(0.33+0.022\xi_{m-1,0})}\right] \quad (17)$$

meanwhile Eq.(4) can be now expressed as:

$$\frac{q}{\sqrt{gH_{m0,t}^3}} = \gamma_\sigma \exp\left[-\left(A + B\frac{R_c}{H_{m0,t}}\right)\right] \qquad (18)$$

Where $\gamma_\sigma$ is expressed by Eqs. (15) and (16).

The comparison between measured mean overtopping discharge and predicted one is depicted in Figure 10 for Eq.(9) and Eq.(17). The left plot shows the measured and calculated discharge using Eq. (9) with no reduction factors. Overtopping of short-crested waves is overestimated when the reduction factor is not applied. Values of wave overtopping discharge around 1 l/s/m, often assumed as threshold for structural design, are better predicted when the reduction factor for short-crestedness is applied.

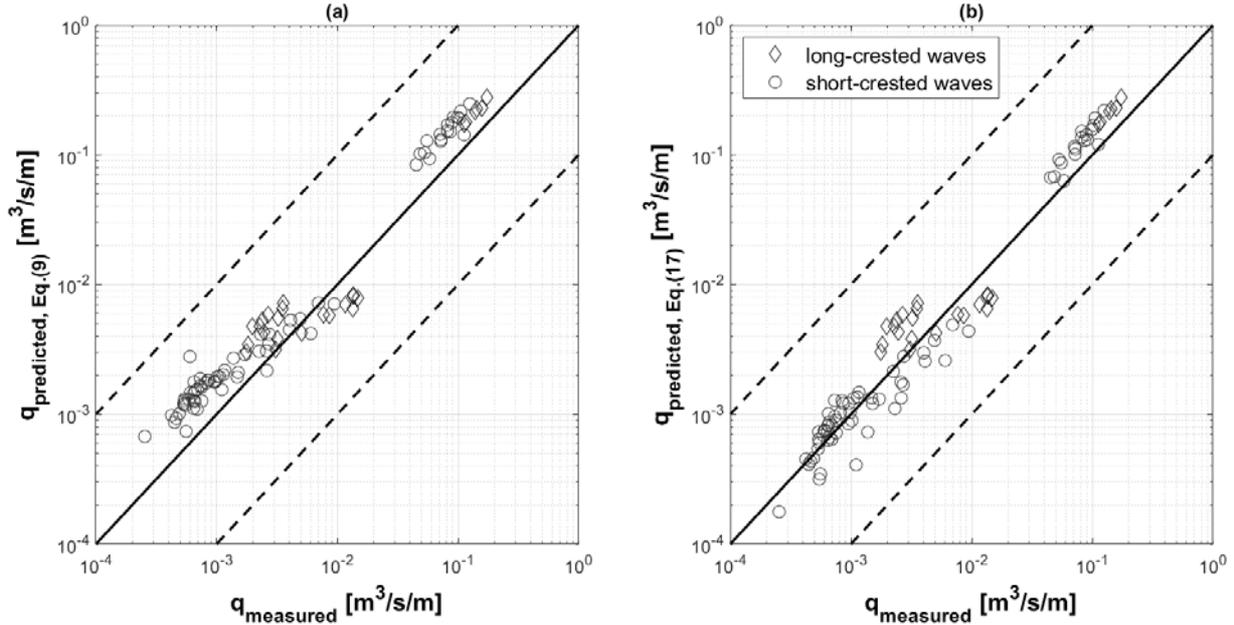

*Figure 10. Comparison between measured overtopping discharge and calculated discharge using Eq.(9) (a) and Eq. (17) with the application of the reduction factors for directional spreading (b).*

When Goda (2009) is applied – see Figure 11a - the results are good for high water level (i.e. very shallow water conditions). However, for low water level (i.e. extreme shallow water conditions) all results start to diverge significantly: Eq. (4) over predicts the mean discharge of about 10 times. This is due to the inaccuracies in the calculation of the coefficients A and B, the expression of which is extrapolated for very low values of $h_t/H_{m0,t}$. In any case, it is evident also from Figure 11 that short-crested wave cases require a further correction in the formula in order to be properly predicted. The right plot Figure 11 show the measured discharge versus the one predicted using Eq. (18). The prediction improves significantly when reduction factors are applied.

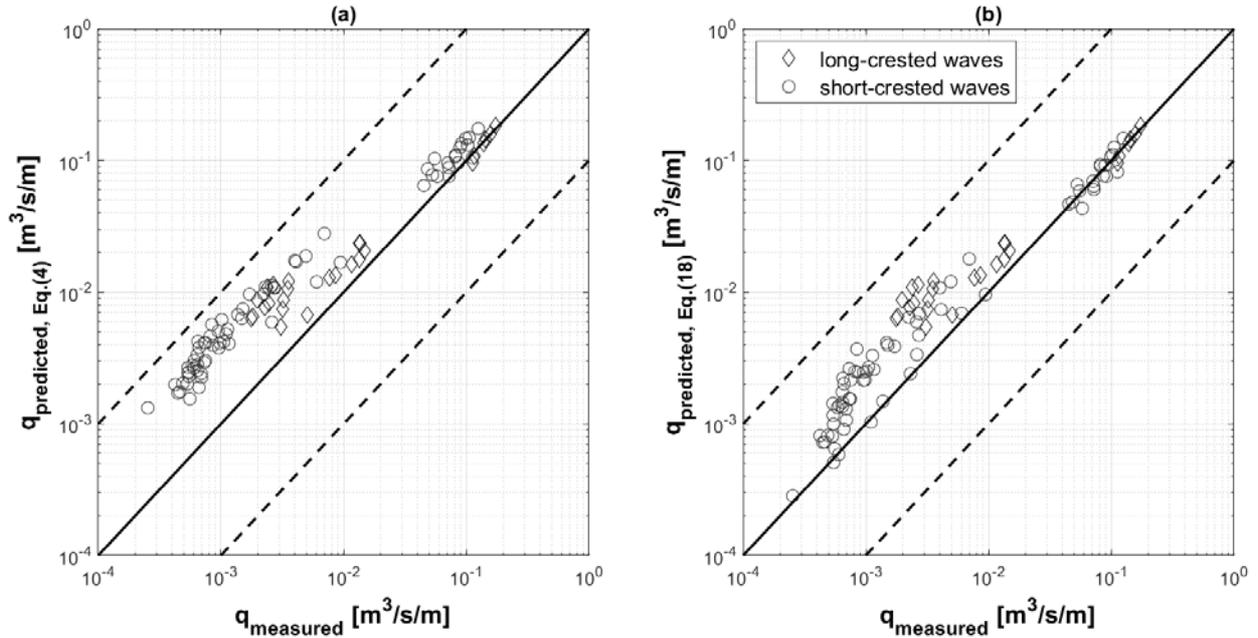

*Figure 11. Comparison between measured overtopping discharge and calculated discharge using Goda (2009), without (left picture) and with (right picture) the application of the reduction factors for wave obliqueness and directional spreading.*

The accuracy of the prediction has been evaluated by means of the error estimator proposed in Goda (2009). The geometric mean is expressed as follows:

$$\bar{x}_G = exp\left[\frac{1}{N}\sum_{i=1}^{N} lnx_i\right] \text{ with } x_i = \frac{q_{est,i}}{q_{meas,i}} \quad (19)$$

where N is the number of data, $q_{est,i}$ and $q_{meas,i}$ are the i-th predicted and measured mean overtopping discharge, respectively. The geometric standard deviation is calculated as the exponential value of the standard deviation of the logarithm:

$$GSD = exp\left\{\left[\frac{1}{N}\sum_{i=1}^{N}((lnx_i)^2 - (ln\bar{x}_G)^2)\right]^{0.5}\right\} \quad (20)$$

The geometric standard deviation expresses how much the data are scattered around the geometric mean. Considering a quantity normally distributed, 90% of the data will be contained in the range between $\bar{x}_G/(1.64 \times GSD)$ and $\bar{x}_G \times 1.64 \times GSD$. Table 4 lists all results for geometric mean and standard deviation for the aforementioned formulas, together with the extremes of the range prediction assuming the overtopping normally distributed and the 90% confidence interval. The errors are shown for the two foreshore shallowness regimes, namely $0.3 < h_t/H_{m0,o} < 1$ and $h_t/H_{m0,o} < 0.3$. The application of the reduction factors improves globally the prediction, giving minor errors and narrower prediction intervals.

Table 4. Error calculation: geometric mean and standard deviation

| Formula | Foreshore shallowness | No reduction factors | | | With reduction factor | | |
|---|---|---|---|---|---|---|---|
| | | $\bar{x}_G$ | GSD | 90% $q_{est}/q_{meas}$ interval prediction | $\bar{x}_G$ | GSD | 90% $q_{est}/q_{meas}$ interval prediction |
| van Gent (1999), as modified by Altomare et al. (2016) | $0.3 < \frac{h_t}{H_{m0,o}} < 1$ | 1.63 | 1.57 | 0.63-4.20 | 1.13 | 1.55 | 0.45-2.89 |
| | $\frac{h_t}{H_{m0,o}} < 0.3$ | 1.59 | 1.57 | 0.62-4.08 | 0.99 | 1.54 | 0.39-2.50 |
| Goda (2009) | $0.3 < \frac{h_t}{H_{m0,o}} < 1$ | 1.24 | 1.22 | 0.62-2.50 | 0.99 | 1.15 | 0.52-1.86 |

| | $\frac{h_t}{H_{m0,o}} < 0.3$ | 3.62 | 1.50 | 1.47-8.91 | 2.10 | 1.49 | 0.86-5.14 |

## 6  CONCLUSIONS

The influence of directional spreading on overtopping of sea dike with gentle foreshore in very and extremely shallow water conditions is analysed in the present work. Physical model tests were carried out in the multi-directional wave basin of Flanders Hydraulics Research. Main focus was to characterise the wave transformation and wave overtopping on sea dikes and compare long- and short-crested wave cases. The results show a clear influence of the directional spreading (i.e. wave short-crestedness) on wave transformation and, hence, on wave overtopping: wave height at the dike toe is smaller, wave period shorter and resulting wave overtopping discharge is reduced by almost one order of magnitude if compared to long-crested wave cases. A preliminary analysis on wave order for generation was carried out, while the generation of second-order short-crested waves was not embedded in the generation software. The analysis showed that the differences in overtopping using second- or first-order generation fall within the uncertainties related to the seed number. Therefore, the effect of wave order was neglected in this investigation. However, it is recommended not to generalize this assumption outside the conducted experimental campaign and tested conditions.

The effect of directional spreading has been analysed only for tests with main direction perpendicular to the dike crest. A few tests with oblique wave conditions have been simulated, but only for long-crested waves, due to limitations of the wave generator. Therefore, these tests have been excluded from the analysis and the combined effect of obliqueness and short-crestedness is not taken into account and would require further studies. A detailed analysis of the short-crested wave tests has allowed defining new expressions for the reduction factor, $\gamma_\sigma$, for directional spreading that despite the large data scatter from which is derived, proves to improve significantly the overtopping prediction in the tested range of wave conditions. Yet, considering the particular model layout and the relative small model scale employed for this experimental campaign, namely 1:50, it is recommended that the new factors for wave spreading are employed within the range of tested conditions and for very similar layouts. Besides, it is advised to perform experimental campaign at larger model scale for a better assessment of possible scale effects on the studied phenomenon.

**ACKNOWLEDGEMENTS AND DISCLAIMER**

This research is part of the CREST project funded by the Flemish Agency for Innovation by Science and Technology. Dr. Corrado Altomare, after being involved in CREST till the end of 2018, has received funding from the European Union's Horizon 2020 research and innovation programme under the Marie Sklodowska-Curie grant agreement No 792370.

The presented results reflect only the authors view and the Research Executive Agency (REA) is not responsible for any use that may be made of the information it contains.

**LIST OF FIGURES**

# LIST OF TABLES